\documentclass{ws-procs9x6-cpt19}

\begin{document}

\newcommand{\refeq}[1]{(\ref{#1})}
\def\etal {{\it et al.}}

\title{Status of the KATRIN Neutrino Mass Experiment}

\author{Y.-R.\ Yen}

\address{Department of Physics, Carnegie Mellon University,\\
Pittsburgh, PA 15213, USA}

\author{On behalf of the KATRIN Collaboration}

\begin{abstract}
The  KArlsruhe TRItium Neutrino experiment,
or KATRIN,
is designed to measure the tritium $\beta$-decay spectrum 
with enough precision to be sensitive to the neutrino mass 
down to $0.2\,$eV at 90$\%$ confidence level. 
After an initial first tritium run in the summer of 2018, 
KATRIN is taking tritium data in 2019 
that should lead to a first neutrino mass result.
The $\beta$ spectral shape of the tritium decay 
is also sensitive to four {\it countershaded} Lorentz-violating, 
oscillation-free operators within the Standard-Model Extension 
that may be quite large. 
The status and outlook of KATRIN to produce physics results, 
including Lorentz-violation measurements, 
are discussed.
\end{abstract}

\bodymatter

\section{Introduction}
The massive nature of neutrinos 
remains one of the major open questions in physics 
that cannot be explained by the minimal Standard Model. 
Among the various ways for directly probing the absolute neutrino mass, 
$\beta$ decay is the least model dependent 
unlike cosmology (cosmological models)\cite{cosmology} 
or neutrinoless double beta decay (nuclear models and quenching).\cite{bb0n} Kinematics alone defines the $\beta$-decay spectrum, 
from which the electron-antineutrino mass, 
in the form of the effective mass 
$m^2_{\mathrm{eff}}(\nu_{\mathrm{e}}) = \Sigma_\mathrm{i} |U^2_{\mathrm{ei}}|m^2_{\mathrm{i}}$, will determine the spectral-shape modification 
very near the high-energy endpoint. 

A good candidate for very precise $\beta$ spectral measurements is tritium (T). Tritium ($^3$H) decays into $^3$He 
with the emissions of an electron and an antineutrino. 
The relatively low $Q$ value of $18.3\,$keV means 
that the fractional change to the spectrum 
by the neutrino mass would be large. 
A high-luminosity tritium source, 
necessary for high statistics 
since only a small fraction of the decays are near the endpoint energy, 
can be designed due to the tritium half-life being 12.3 years.

The previous generation of tritium beta-decay experiments, 
located at Troitsk and at Mainz, 
set a limit on the antineutrino mass at $\sim\!2\,$eV at 90$\%$ c.l.\cite{Troitsk, Mainz}
KATRIN is a next-generation tritium experiment 
that has recently started taking data. 
The ultimate sensitivity of KATRIN 
is expected to be $\sim\!0.2\,$eV at 90$\%$ c.l., 
an improvement by an order of magnitude.

\section{Experiment Description}\label{experiment}
The KATRIN experiment can achieve its desired sensitivity 
due to a high-activity, gaseous molecular-tritum (T$_2$) source 
and a high-resolution spectrometer. 
A $\beta$ electron will be adiabatically guided by magnetic fields 
to travel down the entire 70-meter beamline to the detector. 
After the decay takes place in the windowless, gaseous T$_2$ source, 
a two-stage transport section eliminates the non-$\beta$-electron components 
(ions and neutral atoms) ahead of the main spectrometer. 
The main spectrometer, 
with the Magnetic Adiabatic Collimation 
combined with an Electrostatic 
filter design, 
will select with fine energy resolution 
($0.93\,$eV at the endpoint energy of $18.6\,$keV) 
only $\beta$ electrons above a specific energy to pass. 
This filter adiabatically transforms the $\beta$ electrons' momenta 
to the longitudinal direction, 
the same direction as the precisely set spectrometer voltage, 
allowing the endpoint silicon PIN diode detector 
to count essentially just the number of electrons 
with decay energies above an energy threshold. 
A more detailed description of the KATRIN experimental setup 
can be found in Ref.\ \refcite{KATRIN_JINST}. 
In normal KATRIN operation, 
an integrated spectrum of $\beta$ decay is measured. 

The measurement-time distribution, 
i.e.,
where and how long to scan the $\beta$ spectrum 
via the main spectrometer potential setting, 
is optimized for the best neutrino mass sensitivity. 
Reference \refcite{KATRIN} lists the theoretical modification to the $\beta$ spectrum 
that KATRIN will need to consider. 
From the integrated spectrum with those effects considered, 
the neutrino mass will be one of four fundamental parameters to be fitted, 
alongside the background, 
the amplitude,
and the endpoint energy,
which is also important for Lorentz-violation (LV) searches.

\section{Prospects for LV searches with KATRIN}
The Standard-Model Extension (SME) provides a framework 
for small, testable Lorentz and CPT breakdown.\cite{SME} 
In the neutrino sector of the SME, 
the majority of LV coefficients are expected to be heavily suppressed 
by the tiny ratio of the electroweak and Planck scales 
($m_w$/$m_p$) of about $10^{-17}$. 
However, 
four oscillation-free operators 
may be examples of the rare {\it countershaded} LV; 
these effects may be large but difficult to detect, 
akin to a shark's different color on its belly and its back.  
These four operators are renormalizable, 
flavor blind, 
and of mass dimension three; 
other than breaking Lorentz and CPT symmetry, 
all other physics is conventional.\cite{LV_betadecay}

The corresponding countershaded coefficients to these operators 
are one timelike $(a^{(3)}_{\mathrm{of}})_{00}$ 
and three spacelike $(a^{(3)}_{\mathrm{of}})_{1m}$ 
with $m = 0,\pm 1$. 
While the $00$ term is independent of apparatus location, 
the spacelike terms depend on $\chi$, $\xi$, and $\theta_0$ 
representing the coordinate change 
to the standard choice of SME frame;
for terrestrial experiments like KATRIN, 
the leading contribution to this transformation 
arises from the rotation of the Earth around its axis.

The paper by 
D\'\i az \etal, 
on {\it Lorentz violation and beta decay}\cite{LV_betadecay} 
both analyzed the published Troitsk and Mainz results\cite{Troitsk, Mainz} 
to get the current best limits 
on two of the oscillation-free operators 
and provided the road map to figure out the other two operators, 
which requires binning the data based on sidereal time. 
The rest of this section will mostly summarize that paper. 

For the tritium-decay energy range $\Delta T$ near the endpoint energy $T_0$,
i.e.,
$\Delta T = T_0 - T$ is small, 
the decay rate can be written as
\begin{equation}
\frac{d\Gamma}{dT} 
\simeq B + C\Big[\big(\Delta T + k(T_{\oplus})\big)^2-\tfrac{1}{2}m_{\nu}^2\Big],
\label{aba:eq1}
\end{equation}
where $B$ is the experimental background rate and $C$ is approximately constant. 
The SME coefficients contribute to the function $k(T_{\oplus})$,
which depends on the sidereal time $T_{\oplus}$:
\begin{eqnarray}
k(T_{\oplus}) &=& 
\tfrac{1}{\sqrt{4\pi}}(a^{(3)}_{\mathrm{of}})_{00}
	-\sqrt{\tfrac{3}{4\pi}} \cos^2\!\tfrac{1}{2}\theta_0\,
	\sin{\chi}\,\cos\xi\,(a^{(3)}_{\mathrm{of}})_{10} 
\nonumber\\
	&&{}-\sqrt{\tfrac{3}{2\pi}}\cos^2\!\tfrac{1}{2}\theta_0\,
	\Big[\sin\xi\,{\mathrm{Im}}
	\big((a^{(3)}_{\mathrm{of}})_{11}e^{i\omega_{\oplus}T_{\oplus}}\big)
	\nonumber\\
	&&{}+\cos{\chi}\,\cos\xi\,{\mathrm{Re}}\big((a^{(3)}_{\mathrm{of}})_{11}^*e^{-i\omega_{\oplus}T_{\oplus}}\big)\Big],
\label{aba:eq2}
\end{eqnarray}
where ${\mathrm{Re}}(a^{(3)}_{\mathrm{of}})_{11} = (a^{(3)}_{\mathrm{of}})_{11}$ and ${\mathrm{Im}}(a^{(3)}_{\mathrm{of}})_{11} = (a^{(3)}_{\mathrm{of}})_{1-1} \equiv -(a^{(3)}_{\mathrm{of}})_{11}^*$.

Equation \ref{aba:eq2} shows 
that $k(T_{\oplus})$ will only shift the endpoint energy of the decay spectrum 
without changing the shape 
and is independent of the neutrino mass. 
Without knowing the sidereal time, 
the harmonic oscillation effects of the $1-1$ and $11$ terms on the spectrum 
will average out. 
Limits on $({a}^{(3)}_{\rm of})_{00}$ and $({a}^{(3)}_{\rm of})_{10}$ 
can then be obtained 
from the limit on the potential endpoint energy shift.

By conservatively taking $<5\,$eV to be the Troitsk and Mainz sensitivity, 
$|(a^{(3)}_{\mathrm{of}})_{00}| < 2 \times 10^{-8}\,$GeV 
and $|(a^{(3)}_{\mathrm{of}})_{10}| < 5 \times 10^{-8}\,$GeV limits were set. 
For comparison, 
the neutrinoless double beta-decay experiment EXO-200 
also searched for $(a^{(3)}_{\mathrm{of}})_{00}$ 
from the two-neutrino double beta-decay spectrum shape;\cite{EXO200_LV} 
their limit of 
$-9.39 \times 10^{-5}\,$GeV$ <(a^{(3)}_{\mathrm{of}})_{00} 
< 2.69 \times 10^{-5}\,$GeV 
is significantly worse 
than what a single $\beta$-decay experiment can do.

Reference \refcite{LV_betadecay} predicts 
that 30 days of nominal KATRIN run 
can improve the limit on $(a^{(3)}_{\mathrm{of}})_{00}$ 
by two orders of magnitude.
A KATRIN analysis that considers sidereal time 
can set the first limits 
on $(a^{(3)}_{\mathrm{of}})_{11}$ and $(a^{(3)}_{\mathrm{of}})_{1-1}$.

\section{Outlook}
An initial tritium injection June 5th--18th, 2018 
resulted in a successful ``First Tritium" run of 81 hours worth of tritium data. 
The data taken from March until May of 2019 
with a higher concentration of tritium, 
named ``KNM1," 
will be the first KATRIN run to be used for a neutrino mass result. 
With some off-season adjustments, 
KATRIN plans to operate at the nominal settings starting in 2020. 
A three-year run spanning five calendar years 
should allow KATRIN to reach the design sensitivity of $0.2\,$eV.

The timeline for a KATRIN LV result 
will depend on the prior release of neutrino mass results. 
Even if the blinding scheme for the neutrino mass analysis 
does not mask the endpoint energy result 
that is needed for the LV analysis, 
a comprehensive understanding of the endpoint energy systematics 
will not be available until after the completion of the main neutrino mass analysis. 
With some strong theoretical motivations to search for LV 
with the $\beta$-spectrum method, 
KATRIN will likely produce LV results 
following each neutrino mass result release 
in the coming years.

\section*{Acknowledgments}
This work is supported by the DOE Office of Science under award no.\ $\#$DE-SC0019304.


\begin{thebibliography}{x}

\bibitem{cosmology}
Planck Collaboration,
N.\ Aghanim \etal,
arXiv:1807.06209.

\bibitem{bb0n}
M.J.\ Dolinski, A.W.P.\ Poon, and W.\ Rodejohann,
arXiv: 1902.04097.

\bibitem{Mainz}
C.\ Kraus \etal,
Eur.\ Phys.\ J.\ C {\bf 40}, 447 (2015).

\bibitem{Troitsk}
V.N.\ Assev \etal,
Phys.\ Rev.\ D {\bf 84}, 112003 (2011).

\bibitem{KATRIN_JINST}
KATRIN Collaboration,
M.\ Arenz \etal,  
JINST {\bf 13}, P04020 (2018).

\bibitem{KATRIN}
M.\ Kleesiek \etal, 
Eur.\ Phys.\ J.\ C {\bf 79}, 204 (2019). 

\bibitem{SME}
D.\ Colladay and V.A.\ Kosteleck\'y, 
Phys.\ Rev.\ D {\bf 55}, 6760 (1997);
D.\ Colladay and V.A.\ Kosteleck\'y, 
Phys.\ Rev.\ D {\bf 58}, 116002 (1998);
V.A.\ Kosteleck\'y, 
Phys.\ Rev.\ D {\bf 69}, 105009 (2004).

\bibitem{LV_betadecay}
J.S.\ D\'\i az , V.A.\ Kosteleck\'y , and R.\ Lehnert,
Phys.\ Rev.\ D {\bf 88}, 071902 (2013).

\bibitem{EXO200_LV}
EXO-200 Collaboration,
J.B.\ Albert \etal, 
Phys.\ Rev.\ D {\bf 93}, 072001 (2016).

\end{thebibliography}
\end{document}